\documentclass{styles/svproc}

\usepackage{algorithm}
\usepackage{algpseudocode}
\usepackage[numbers,sort&compress]{natbib}
\usepackage{amsmath}
\usepackage{hyperref}

\usepackage{subcaption}
\usepackage{wrapfig}
\usepackage{graphicx}
\usepackage[stable]{footmisc}

\usepackage{url}

\begin{document}
\mainmatter              % start of a contribution
\title{Testing the validity of multiple opinion dynamics models}
\titlerunning{Opinion Dynamics}  % abbreviated title (for running head)
% \author{Samuel Moor-Smith\inst{1[0000-0002-6493-5994]} \and \\Dino Carpentras\inst{1[0000-0001-8471-2352]}}
\author{Samuel Moor-Smith\inst{1} \and Dino Carpentras\inst{1}}
\authorrunning{Samuel Moor-Smith et al.} % abbreviated author list (for running head)
% %
% %%%% list of authors for the TOC (use if author list has to be modified)
% % \tocauthor{Dino Carpentras \and Samuel Moor-Smith}
% %
\institute{ETH Zürich, 8092, Zürich, Switzerland\\
\email{samuelmoorsmith@gmail.com},\\ WWW home page:
\texttt{https://coss.ethz.ch/}}

\maketitle              % typeset the title of the contribution

\begin{abstract}
While opinion dynamics models have been extensively studied as stylized models, there has been growing attention to the possibility of combining these models with empirical data. This attention seems to be driven by the many social issues that strongly depend on people's opinions (such as climate change and vaccination) and the need for empirically valid models to design related policy interventions. While different models have been combined in various ways with empirical data, standardised comparison of models against empirical data is still lacking.
In this article, we test the validity of multiple opinion dynamics models---including both stylized and more realistic models. Our approach follows a ``data science-like" validation procedure, where we first calibrate the model's free parameters using an initial range of years (e.g. 2010-2015), and then use data from one wave (e.g. 2016) to predict data in the following wave (e.g. 2017).
We initially tested such a procedure using simulated data and then tested different models on various topics from the European Social Survey. Both toy models and empirical models perform well on the simulated data, but fail to predict future years in the empirical data. Furthermore, during the calibration phase on the empirical data, most models learned to ``freeze"---meaning that their predictions for the following year are just a copy of the data from the previous year.
This work advances the literature by offering a benchmark for comparing different opinion dynamics models. Furthermore, our tests show that real-world dynamics appear to be completely incompatible with the dynamics of the tested models. This calls for more effort in exploring what are the features that would improve validity and applications for opinion dynamics models. 
% We would like to encourage you to list your keywords within
% the abstract section using the \keywords{...} command.
\end{abstract}
\section{Introduction}
For a long time, opinion dynamics has focused on stylized models that are able to show some interesting properties of opinion interactions~\citep{flache2017models, castellano2012social, xia2011opinion}. For example, models like the Deffuant model~\citep{deffuant} or the Axelrod model~\citep{axelrod1997dissemination} have shown how simple interactions can lead to the formation of coherent opinion-based groups, similarly to how the Schelling model has shown the appearance of segregation patterns~\citep{schelling1971dynamic}. These models have driven the field of opinion dynamics and provided many valid interpretations of how opinions form, shift, and evolve based on theories in psychology and sociology.

However, in recent years, societies have started facing multiple pressing problems related to public opinion. For example, climate change can be effectively addressed only if people agree on what to do~\citep{climate_change_coordination}, vaccines can produce herd immunity only if adopted by most people~\citep{vaccination_coverage}, and countries can face instability if people's ideological divide becomes too strong~\citep{polarization}. Many of these challenges call for well-informed public policy, which opinion-dynamics models can help to support~\citep{edmonds2023grand}. 

This has driven further research into understanding the more realistic features in stylized models~\citep{geschke2019triple, grabisch2023design}. Others have attempted a more empirical approach by deriving models from experiments, or testing if their models could reproduce some empirical patterns~\citep{gestefeld2023, banisch2024, carpentras2023psychometric, Devia2022, Monti2023}.

While all these approaches explore some kind of model validity, they present at least one of the following three main limitations: \textbf{(1)} data and methods are often very specific and do not generalize to other common datasets; \textbf{(2)} it is often not clear if the model is valid or not; and \textbf{(3)} temporal predictions are often neglected.

The first problem refers to the fact that most studies rely on data and methodologies that are very context- or model-specific, and make it hard to use a similar methodology in other studies. For example,~\citep{carpentras2022mapping} used opinion data related to vaccination in 2018 to predict vaccination coverage in 2019. While this is an interesting implementation, it is unclear how this could be used for other datasets and years. In another example,~\citep{banisch2024} et al. used macro-level data to estimate a micro-level parameter, a methodology that requires a very specific data collection process.

The second problem refers to the fact that results are often given as pure numbers. For example, we may estimate that a model, $M_1$, produces an estimation error of 5.0, while model $M_2$ produces an error of 4.8. While everybody can agree that 4.8 is an improvement on 5.0, it is not clear if this improvement is significant in any way. For example, simply presenting the results in this way, it may seem as though $M_2$ performs well on empirical data. However, we may discover that a purely random model produces an error of 2.3, thus even if $M_2$ is an improvement to $M_1$ it still performs so poorly that it should not be used for any empirical application.

The third and final problem relates to the predictive ability of these models. There is a general split in the social simulation community regarding whether the purpose of such models is primarily explanatory or predictive~\citep{explain_or_predict}. In this paper, we adopt an empirical perspective and therefore consider predictive performance as a key criterion for assessing model validity~\citep{elsenbroich2023agent, chattoe2023agent, dignum2023should}. This does not mean that the model must be deterministic, as we can still have valid and invalid models for probabilistic processes. For example, for a fair dice, a valid model would be one that predicts 1 to 6 with equal probability, while supposing 1 as the unique possible result would be an invalid model. Similarly, an opinion dynamics model that often makes bad predictions\footnote{By ``prediction" here we mean the ability of a model to estimate data that are in the future respect to the input data. For example, using 2010 data to ``predict" 2015 data. Thus, our use of prediction is only related to the temporal positioning of the data and has nothing to do with ``the future" with respect to the time of writing.} should not be used for policy applications. Despite this key role of prediction, they are rarely used for testing opinion dynamics models' validity.

For all these reasons, the remainder of the article presents a methodology for testing multiple opinion dynamics models. Specifically, Section 2 outlines our approach to model calibration and validation; Section 3 evaluates each model’s performance in predicting and reconstructing synthetic opinion data; and Section 4 applies the models to empirical data.

\section{Methodology for Model Calibration and Validation}

% In this section, we outline the methods used to calibrate and validate the opinion dynamics models. We describe the chosen loss function and its role in quantifying the differences between two opinion distributions. Additionally, we introduce the explained variance as a key metric for assessing model performance. This metric evaluates how well a model captures opinion shifts over time by comparing its predictions to those of a "No-Change" (null) model---a baseline that assumes opinions remain static from one time step to the next. A model that significantly outperforms the null model demonstrates its ability to capture meaningful opinion dynamics. Finally, we introduce the selected optimization algorithm and the general evaluation framework, which are applied consistently across all experiments in this study.

In this section, we describe the terminology and methods used to calibrate and validate the opinion dynamics models, including the loss function, the explained variance metric, the optimization algorithm, and the general evaluation framework applied across all experiments.

\subsection{Terminology and notation}
% In the next sections, we will discuss  how to compare multiple datasets and how to run opinion dynamic models on them. In order to improve clarity, here we define the terms and operations that we will use.

\subsubsection{Datasets:} We refer to a dataset $D$ as a collection of opinion distributions $O_t$, each one related to a time $t$. For example, a dataset containing opinion distributions annually collected from 2002 to 2018 will be represented as  $D = \{ O_t \}_{t=2002}^{2018}$.

\subsubsection{Models as operators:} For our purposes, an opinion dynamics model $M$ is an operator that takes in an opinion distribution $O_t$ as an input and produces a new opinion distribution $\hat{O}_{t+1}$, where $t$ represents a wave of the dataset\footnote{Note that consecutive waves may not be in consecutive years. For example, data may be collected every 2 years, and so the wave after 2016 will be 2018.}. For example, if we are applying model $M$ on 2016 data to predict 2018 data, we will write $M(O_{2016})=\hat{O}_{2018}$. This notation will become useful when we will compare the ``original" 2018 data (i.e. $O_{2018}$) with the one predicted from the model (i.e. $\hat{O}_{2018}$). Furthermore, we will refer to the original data as the \emph{ground-truth}\footnote{While most empirical datasets on opinions, attitudes, or values are cross-sectional, meaning that different individuals are surveyed at each time point, we assume that with a sufficiently large sample size the aggregate (macro-level) characteristics remain representative of the underlying population. The models are designed to capture population-level dynamics, so this assumption allows us to use the available cross-sectional data as a practical---albeit imperfect---proxy for longitudinal ground-truth.}. This term will be used both when the original data are empirical and when they have been obtained from other simulations.

% Since opinion dynamics models are often random, in some cases, it makes sense to apply the same data to obtain multiple outputs. When we repeat n-times the application model M on the input distribution $O_{t-1}$ we write:
% \begin{equation}
%     M^{n}[O_{t-1}] = [\hat{O}^1_t,\hat{O}^2_t, ... \hat{O}^n_t]
% \end{equation}
% So each  $\hat{O}^i_t$ is a prediction of the same time point, with the same model and same input data. The only difference is the random interactions used by the model.

\subsubsection{Prediction Error:} Given a certain model $M$ we define its prediction error for a certain dataset (consisting of T opinion distributions) as:
\begin{equation}
    E(M,D) = \sum_{t=2}^T \Delta(O_t, \hat{O}_t)
\end{equation}
where $\Delta$ is the Wasserstein distance between the two distributions~\citep{wasserstein}. The Wasserstein metric implementation we use is from \texttt{scipy}~\citep{scipy}. $\hat{O}_t$ is the models' predicted value for timestep, $t$: $\hat{O}_t = M(O_{t-1})$. 

% Notice that we will always use one wave to predict the following one. However, please notice that consecutive waves may not be in consecutive years. For example, data may be collected every 2 years, and so the wave after 2016 will be 2018.

\subsubsection{Explained Variance and Opinion Drift:} While the error can tell us which models perform better or worse, it cannot directly tell us if a model is performing well or not. Indeed, we know that an error of 1.1 is better than an error of 1.2, but we cannot really tell if this is an excellent or very poor result. To avoid this problem, we introduce a null (``No-Change") model. Such a model, called $M_0$, predicts that the following distribution will simply be identical to the previous distribution, so that:
\begin{equation}
  M_0(O_t) = O_t
\end{equation}
Thus the prediction error of the null model will always be equal to the variance in the data between all of the years:
\begin{equation}
    E(M_0,D) = \sum_{t=2}^T \Delta(O_t, O_{t-1})
\end{equation}
Since $E(M_0,D)$ relies solely on the dataset and also captures the amount of opinion change from one timestep to the next, we can define a new term:
\begin{equation}
    \Delta_{\text{op}}(D) = E(M_0,D) = \sum_{t=2}^T \Delta(O_t, O_{t-1})
\end{equation}
We will call $\Delta_{\text{op}}(D)$ the \textbf{opinion drift of dataset $D$} and this will be mentioned extensively in this study. Note that a low $\Delta_{\text{op}}(D)$ implies that there is little opinion change in the dataset across timesteps, and a high $\Delta_{\text{op}}(D)$ implies that there is a large amount of opinion change across timesteps.

We can now use this null model to produce a normalized measurement of a model $M$'s prediction using the formula of the explained variance V:
\begin{equation}
    V(M,D) = 1- \frac{E(M,D)}{E(M_0,D)} = 1- \frac{E(M,D)}{\Delta_{\text{op}}(D)}
\end{equation}
Similarly to the classical explained variance, when model a $M$'s predictions are perfect (i.e. $E(M,D)=0$) the explained variance is 1; when the model performs as well as the null model (i.e. $E(M,D)=E(M_0,D)$), the explained variance is 0; and if our model performs worse than the null model, the explained variance is negative.\footnote{The introduction of the opinion drift and, consequently, the normalized explained variance enables us to compare the performance of variance models in a standard way. The null model is an unbiased benchmark to compare the provided model against.}

\subsection{Models and Free Parameters} 

We conduct our evaluation on four opinion dynamics models, comprising two foundational ``toy" models and two empirically motivated models. The toy models include the Deffuant Model~\citep{deffuant} and the Hegselmann–Krause (HK) Model~\citep{Hegselmann2005}, both based on the principle of bounded confidence. %---with the latter, we used an arithmetic mean to update opinions.
The empirical models include the ED (Experimentally-Derived) Model~\citep{carpentras2022}, whose dynamics were derived from experimental data, and the Duggins Model~\citep{Duggins_2017}, which incorporates multiple psychological effects through mechanisms such as intolerance, susceptibility and conformity.

Each opinion dynamics model includes a set of \textbf{free parameters} that can be tuned to control the model's behaviour and the dynamics of opinion evolution. For example, in the Deffuant model~\citep{deffuant}, the parameters include $\mu$ (the convergence parameter) and $\epsilon$ (the bounded confidence threshold). When referring to a specific instance of a model with a given parameter set, $p$, we denote it as $M_p$.

\subsection{Optimization Method}
\label{optimisation_method}

Assume that we have a ground-truth dataset, \( D_{\text{true}} \), which contains the observed opinion distributions at each time step: $D_{\text{true}} = \{O_1, O_2, ..., O_T\}$.

For the free parameter optimization, we only use a subset of \( D_{\text{true}} \), which we denote by \( D_{\text{opt}} \subset D_{\text{true}}\). This subset consists of the first $T_{\text{opt}}$ timesteps of $D_{\text{true}}$:
\begin{equation}
    D_{\text{opt}} = \{O_1, O_2, \ldots, O_{T_{\text{opt}}} \}, \quad \text{ where } T_{\text{opt}} < T
\end{equation}
The remaining timesteps are reserved for model validation.

We optimize the free parameters, $p$, of a model $M_p$ by minimizing its prediction error: $E(M_p,D_{\text{opt}})$. Since the models are stochastic, we run each parameter configuration multiple times and use the average error as the loss for that trial.

%To ensure that this method generalizes to real-world scenarios---where we likely will not have access to the \emph{groundtruth} data---we train the optimizer using only the first half of the timesteps. The entire sequence is then used to evaluate model performance after training.

For the specific optimizer, we chose to use the TPE algorithm~\citep{tpe} from the \texttt{hyperopt} Python library~\citep{hyperopt}. This optimizer was previously employed by Duggins~\citep{Duggins_2017} for calibrating opinion dynamics models, and has proven to be effective in a variety of other domains~\citep{hpopt_comparison}.

% For each parameter set we test, we generate 5 runs and use the average loss of these runs as the recorded loss for that parameter set. 

\section{Simulated Data: Experiments 1 - 3}

%In this section, we  discuss the experiments conducted using simulated data. We begin by outlining the methodology specific to these experiments, followed by a presentation of the results.

\subsubsection{Generation of the (Simulated) Ground-Truth Data.} 
\label{groundtruth_generation}
In each of the following experiments, we will generate $K$ \emph{simulated} ground-truth datasets. Each of these can be understood as representing the evolution of an opinion distribution on a particular topic over time. Each dataset will be generated as such:
\begin{enumerate}
    \item Let $M$ be the tested model.
    \item Randomly sample a set of \textbf{generator parameters}  $p^*$, and set the free parameters of the model to get $M_{p^*}$. We will refer to this as the \textbf{generator model}.
    \item Randomly sample an initial set of opinions $O_1$ from a uniform distribution.
    \item Run the model $M_{p^*}$ starting from $O_1$ for $\tau$ time steps to obtain $O_2$. Then input $O_2$ and run the model for another $\tau$ steps to obtain $O_3$. This process is repeated until reaching $O_T$.
    \item The resulting sequence of opinion distributions constitutes \textit{one} \textbf{simulated ground-truth dataset}: $D_{\text{true}} = \{ O_t \}_{t=1}^{T}$.
\end{enumerate}

In the following experiments, $K=100$, $T=10$, $T_{\text{opt}}=5$, and the number of agents is set to 1000. For both the Duggins and HK models, $\tau$ is set to 1, since these models update all agents in 1 iteration. For the Deffuant and ED models, $\tau$ is treated as a free parameter that is learned by the optimizer.
%The value of $\tau$ is dependent on the model.\footnote{For both the Duggins and HK models, $\tau$ is simply 1, since these models update all agents in 1 iteration. For the Deffuant  and ED models, $\tau$ is treated as a free parameter that is learned by the optimizer.}
%With the ground-truth data generation method established, we now proceed to our first experiment.

\subsection{Experiment 1: Model's Reproducibility} 
\label{reproducibility}

\subsubsection{Opinion Change Required for Model Reconstruction.}
Suppose we are given a ground-truth dataset $D_{\text{true}}$, generated by an unknown generator model $M$. To reliably reconstruct $M$, the dataset must contain a sufficient degree of opinion change over time. If opinions remain largely stable, there are only weak signals to the optimizer, making it harder to identify the parameters that drive the model’s dynamics.

We hypothesize that the amount of opinion change required for model identification is related to the model’s \textbf{reproducibility}---its ability to produce consistent outcomes from the same initial conditions

To quantify this, we measure the explained variance obtained when re-running the generator model with the same initial distribution and the same parameters used to generate the data. This reflects how closely the model can reproduce its own dynamics when everything is held constant.

We perform the following steps for each of the $K$ ground-truth datasets:
\begin{enumerate}
    \item We use the same generator model, $M_{p^*}$, that was used to generate the dataset: $D_{\text{true}} = \{O_1, O_2, ..., O_T\}$.
    \item We also use the same initial opinion distribution $O_1$ to ensure that the entire initial configuration is the same as in the ground-truth case. Therefore, the only resulting differences will be as a result of model stochasticity.
    \item We run the model $M_{p^*}$ for $T*\tau$ time steps, starting from $O_1$. At each step $t$, we feed the corresponding ground-truth opinion distribution $O_t$ from $D_{\text{true}}$ into the model to generate the next prediction: $M_{p^*}(O_{t}) = \hat{O}_{t+1}$.
    \item Calculate the explained variance, $V(M_{p^*},D_{\text{true}})$, averaging results across multiple runs to account for stochasticity.
%     \item Repeat the above simulation (K=10) times to account for stochasticity. For each run \( k \), compute the explained variance between the predicted and ground-truth distributions: $E(M,D_{\text{true}})^k$.
%     \item Calculate the average explained variance use this as the reproducibility measure:
% \begin{equation}
%     \text{Avg. } E(M,D_{\text{true}}) = \frac{1}{10}\sum_{k=1}^{10} E(M,D_{\text{true}})^k
% \end{equation}
    % \item This explained variance---achieved using the known parameters---will be our reproducibility measure. It represents the upper bound on performance that we expect the optimizer to approach. 
\end{enumerate}

\begin{figure}[h!]
    \centering
    \includegraphics[width=0.72\linewidth]{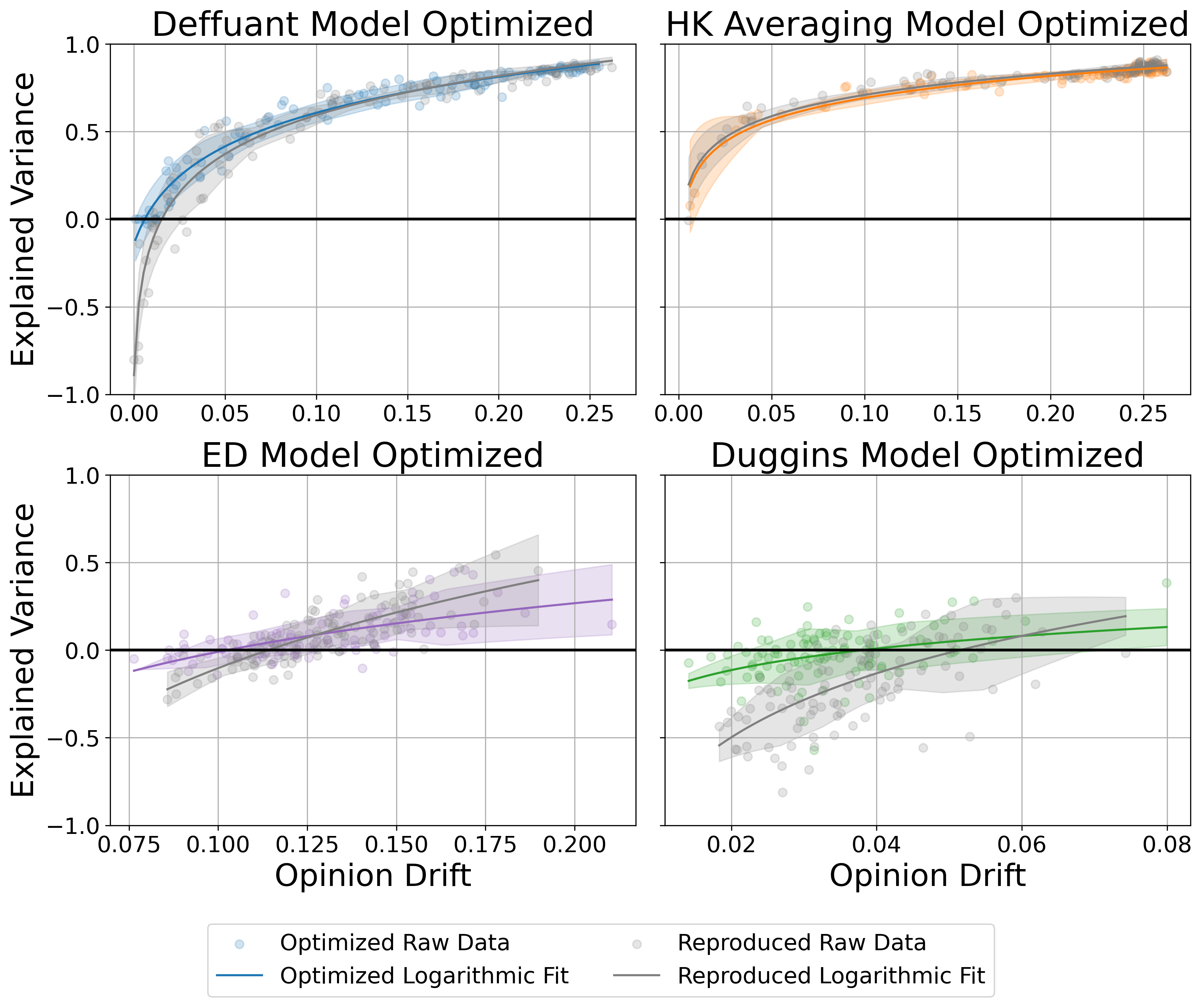}
    % \begin{subfigure}[b]{0.5\textwidth}
    %     \centering
    %     \includegraphics[width=\linewidth]{figures/optimized/combined_optimized.png}
    % \end{subfigure}
    % \begin{subfigure}[b]{0.41\textwidth}
    %     \centering
    %     \includegraphics[width=\textwidth]{figures/deffuant_no_noise.png}
    % \end{subfigure}
    \caption{Explained variance of the optimized model (coloured) and generator model (grey) as a function of $\Delta_{\text{op}}(D)$. For the (grey) ground-truth case (refer. Sec.~\ref{reproducibility}), deterministic models like Deffuant and HK require little opinion drift to outperform the null baseline, while more stochastic models like ED and Duggins require substantially more. The Duggins model shows high variance in repeated runs, reflecting its limited reproducibility. Furthermore, The optimizer matches ground-truth performance at high $\Delta_{\text{op}}(D)$ and often exceeds it at low $\Delta_{\text{op}}(D)$ by mimicking the null model. Shaded areas indicate standard deviation across runs.}
    \label{fig:optimizer_baseline}
\end{figure}

We then map the calculated $V(M_{p^*}, D_{\text{true}})$ for each dataset as a function of the opinion drift: $\Delta_{\text{op}}(D_{\text{true}})$ in Fig. \ref{fig:optimizer_baseline} (grey plots). In this case, high values of the explained variance, $V(M_{p^*}, D_{\text{true}})$, represent the model's ability to reproduce the same dynamics from the same conditions. The opinion drift, $\Delta_{\text{op}}(D_{\text{true}})$, represents the amount of variation in the ground-truth opinions over time, which, in this case, depends on the model, since it has been used to generate the data.

We observe that a minimum amount of opinion drift is required for the explained variance to exceed that of our naive null model baseline, i.e.,
\[
    V(M_{p^*}, D_{\text{true}}) > 0.
\] This threshold varies by model: the HK model requires a negligible amount of opinion change to surpass the baseline, whereas the ED model requires a significantly larger change. We find that the amount of change necessary to outperform the baseline is linked with the stochasticity of the model. The HK and Deffuant models---particularly the HK---are relatively deterministic compared to the two more empirical models. As a result, the amount of change needed to identify model effects is relatively low.

\subsubsection{A Note on the Duggins Model and Reproducibility}

The Duggins model~\citep{Duggins_2017} stood out in our evaluation: even with identical parameters, 72\% of runs performed worse than the naive null model. Two factors explain this:

\textbf{(1)} The parameter ranges used here\footnote{Values: $\mu_t=(0.7,1.0), \mu_s=(1.0,5.0), \mu_c=(0.1,0.5)$, with $\sigma_t, \sigma_c=0.3$, $\sigma_s=0.7$. Grid size: 1000. Social radius: $\mu_r=(0,30)$, $\sigma_r=4$.} match those in~\citep{Duggins_2017}, which were selected to preserve strong diversity. With these parameters, the Duggins model produces datasets with very small amounts of opinion drift. While extending the possible parameter values (e.g., by increasing conformity or intolerance) could increase $\Delta_{\text{op}}(D_{\text{true}})$---and thereby reduce the influence of stochasticity, making the model appear more reproducible---doing so would compromise the integrity of the model.

\textbf{(2)} Unlike the other models, the Duggins model’s ``free parameters" are simply the means and standard deviations of distributions from which each agent’s traits are independently sampled. Combined with the spatial structure of the model, this stochastic resampling introduces significant variability: even when the same parameter values (and agent coordinates) are used, each run resamples each agents parameters, leading to substantially different opinion dynamics. As a result, the model is inherently not reproducible (see the grey plot for the Duggins model in Fig.~\ref{fig:optimizer_baseline}).

\subsubsection{Results of experiment 1} 
These results reveal that, prior to any optimization, the naive null model, $M_0$, can outperform the true generator model, $M_{p^*}$, when the amount of opinion drift is small (i.e. when the dataset presents little variability). A concrete example of this is a scenario is when we have only one pairwise agent interaction between $O_t$ and $O_{t+1}$: the null model will differ only on that one pair, while the generator model---due to stochastic sampling---may not select the correct interaction and will therefore differ both on the true pair and the one it mistakenly simulates. In such cases, the randomness inherent in the generator model can actually worsen its performance relative to a naive baseline.

We also observe that the variance across runs of the generator model differs significantly between models. This reflects the inherent noise in each model's design and the extent to which results from this model can be reproduced. The ED and Duggins models, in particular, exhibit far greater noise than the Deffuant and HK models. Accordingly, a higher level of opinion change is required before a rerun of the generator model performs better than the null model.

These observations are essential for interpreting the results of subsequent experiments. They reinforce the importance of the reproducibility runs, which establish a realistic upper bound on optimizer performance. Since they use the generator model directly, they define the best possible outcome that any optimizer could hope to match under ideal conditions.

% \begin{figure}[h!]
%     \centering
%     \includegraphics[width=0.5\linewidth]{figures/baseline/combined_reproducibility.png}
%     \caption{Each point shows the explained variance achieved by re-running the generator model, averaged over multiple runs. Deterministic models like Deffuant and HK require minimal opinion change to outperform the null baseline, while noisier models like ED and Duggins require substantially more. The Duggins model also shows high variability, reflecting its limited reproducibility.}
%     \label{fig:baseline}
% \end{figure}

\subsection{Experiment 2: Validating the Optimizer}
\label{sec:validating_optimizer}

Having established an upper bound on model performance using known parameters, we now must validate our optimizer. The central question here is whether the optimizer can recover parameters that will result in similar model dynamics.

To do this, we run the optimizer on a subset (i.e. $D_{\text{opt}} \subset D_{\text{true}}$) of the same ground-truth datasets that were generated in Experiment 1\footnote{Recall: if we have $D_{\text{true}} = \{O_1,O_2,...,O_T\}$, we will define $D_{\text{opt}} = \{O_1, O_2, \ldots, T_{\text{opt}}\}$, with $T_{\text{opt}} < T$.}. We use the same model, $M$, and initial opinions, $O_1$, but we \textbf{do not} provide the generator parameters, $p^*$. Instead, we expect the optimizer to find a set of optimized parameters, $\hat{p}$, that result in the same dynamics of opinion. Note that the same dynamics can be achieved by different sets of parameters, so $\hat{p}$
may differ from $p^*$.

Recall that we always feed in the ground-truth data from timestep $O_t$ to predict $\hat{O}_{t+1}$. Further details of the optimization method is described in Section~\ref{optimisation_method}. 

\subsubsection{Optimizer Performance and Model ``Freezing"}
As shown in Fig. \ref{fig:optimizer_baseline}, the optimizer successfully learns parameter values that closely match the explained variance of the generator model across all four opinion dynamics models. In cases where the opinion drift is high, the optimizer performs similarly to the generator model. 

Interestingly, in cases in which the generator model has $V(M_{p^*},D_{\text{true}}) < 0$, we often observe that the optimized parameters, $\hat{p}$, perform better than the generator parameters, $p^*$. Upon inspection of these optimized parameter values, we notice that the optimizer effectively learns to mimic the null model by ``freezing" the opinion dynamics---that is, predicting that opinions remain unchanged between timesteps. This freezing behaviour can be achieved in different ways depending on the model. Examples include: \textbf{(1)} in the Deffuant and ED models, the optimizer can reduce the number of pairwise interactions ($\tau$) close to zero; \textbf{(2)} in the Deffuant and HK models, it can set the bounded confidence threshold ($\epsilon$) near zero, preventing any agents from interacting; and \textbf{(3)} in the Duggins model, the optimizer can reduce the mean social reach ($\mu_r$) to zero, similarly preventing interactions from taking place.

These findings highlight a key behaviour of the optimizer: when the underlying opinion dynamics can be reproduced, it successfully recovers parameter values that can perform competitively to the generator model and the optimizer learns to reproduce the target opinion distributions. Fig.~\ref{fig:noise} illustrates the accuracy of the optimizers fit across nine timesteps. However, in cases where the dynamics are not identifiable---typically due to limited opinion drift---it can still outperform the generator model by suppressing interactions and mimicking the null model.

 %We now investigate its robustness under more realistic settings, where observational noise is introduced.

% \textbf{(right)} Visualization of optimized values fit to the ground-truth data when the ground-truth data is generated from the Deffuant model.

% \begin{wrapfigure}{R}{0.4\textwidth}
%     \centering
%     \includegraphics[width=0.4\textwidth]{figures/deffuant_no_noise.png}
%     % \caption{Simulations and optimization of the Deffuant model. The ground-truth is shown in blue, the predicted distributions obtained from the optimizer in orange.}
%     \caption{}
%     \label{fig:overlap}
% \end{wrapfigure}
% % \begin{figure}[h!]
% %     \centering
% %     \includegraphics[width=0.5\textwidth]{figures/deffuant_no_noise.png}
% %     \caption{Simulations and optimization of the Deffuant model. The ground-truth is shown in blue, the predicted distributions obtained from the optimizer in orange.}
% %     \label{fig:overlap}
% % \end{figure}

\subsection{Experiment 3: Introducing Noise}

We now introduce controlled amounts of noise into our $K$ simulated ground-truth datasets and evaluate how this affects the optimizer's ability to reconstruct the original model dynamics. This simulates if the optimization is successful even when ground-truth is only partially produced by the tested model.

To simulate noisy opinion trajectories, we add Gaussian noise when generating our ground-truth datasets. Specifically, we follow the same procedure as in Sec.~\ref{groundtruth_generation} (\textbf{Generation of the Ground-Truth Data}), except after each opinion distribution $O_t$ is generated, we add gaussian noise to get: $\tilde{O}_t \sim  \mathcal{N}(O_t, \sigma^2)$. We then input $\tilde{O}_t$ to get the next distribution $O_{t+1}$. The noisy dataset will be denoted $\tilde{D}_{\text{true}} = \{O_1, \tilde{O}_2, ..., \tilde{O}_T\}$. The standard deviation $\sigma$ is varied systematically across experiments. Our goal is to identify the threshold beyond which model reconstruction becomes infeasible.

\begin{figure}[h!]
    \centering
    \begin{subfigure}[b]{0.5\textwidth}
        \centering
        \includegraphics[width=\linewidth]{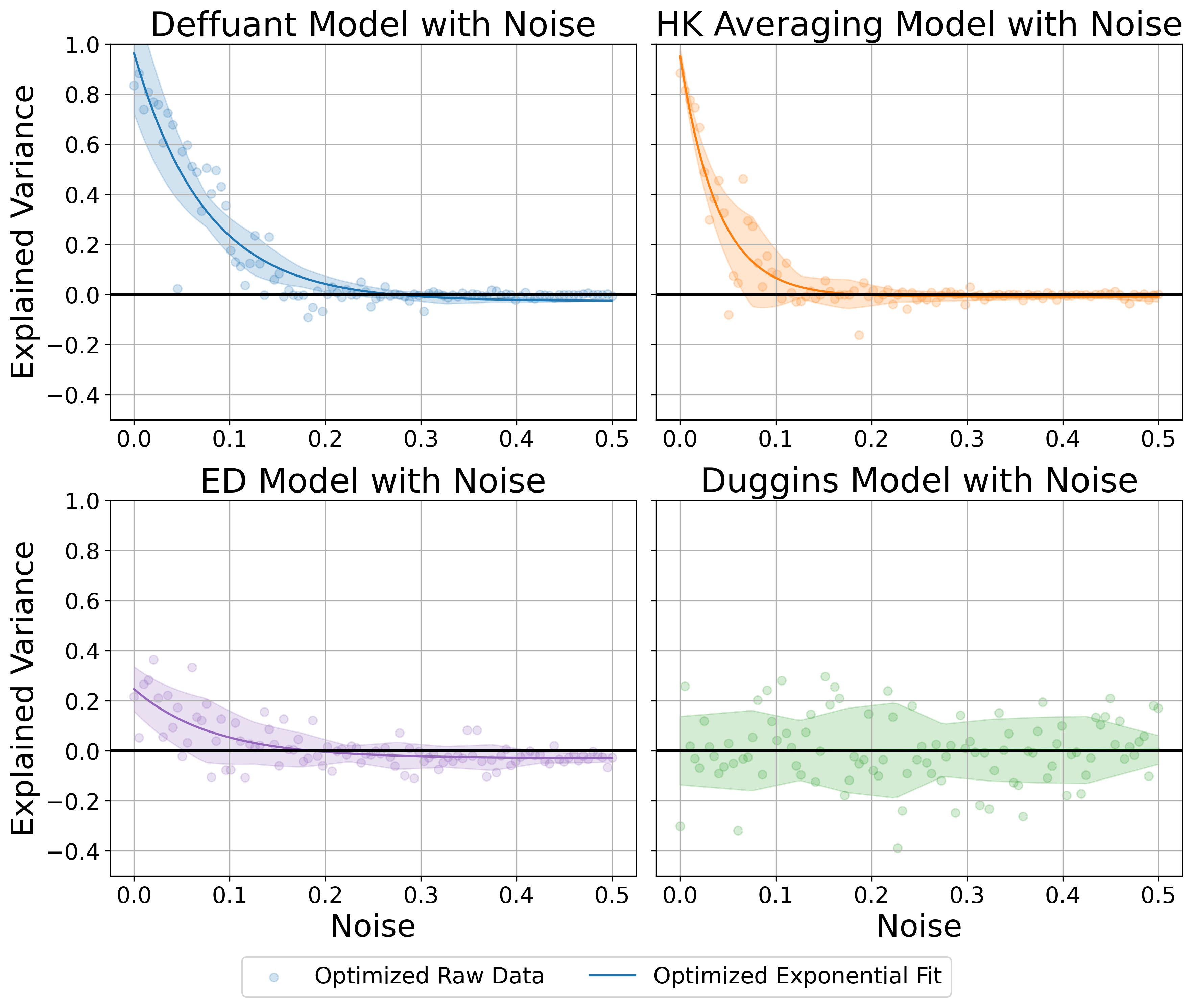}
    \end{subfigure}
    \begin{subfigure}[b]{0.42\textwidth}
        \centering
        \includegraphics[width=\textwidth]{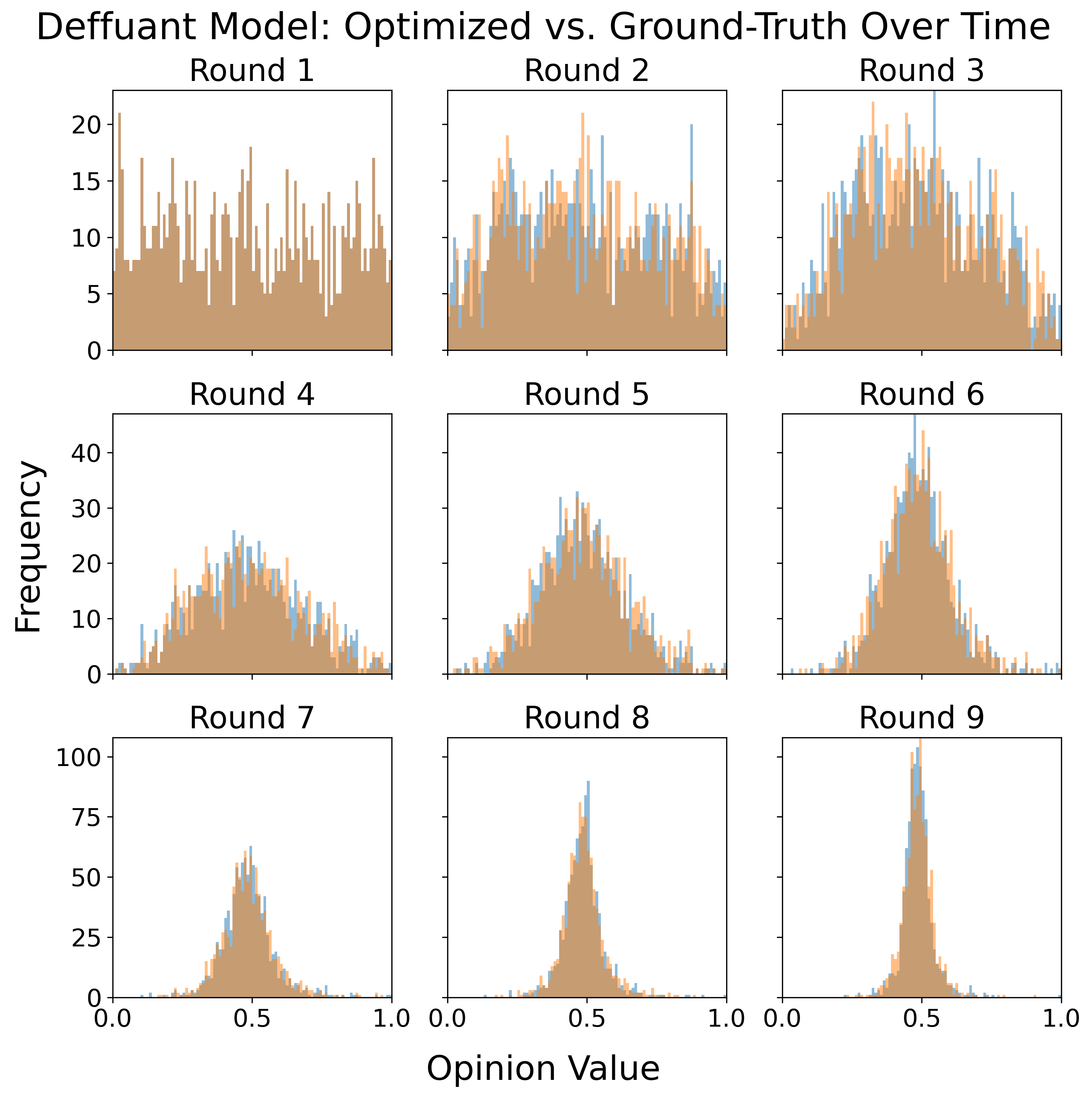}
    \end{subfigure}
    \caption{\textbf{(left)} The optimizer’s performance on a noisy version of a ground-truth dataset. While performance declines exponentially with noise for the Deffuant, HK, and ED models, the Duggins model remains low and variable even without added noise. \textbf{(right)} Visualization of optimized values fit to the ground-truth data when the ground-truth data is generated from the Deffuant model.}
    \label{fig:noise}
\end{figure}

As shown in Fig. \ref{fig:noise}, even small amounts of added noise do not prevent the optimizer from recovering the opinion dynamics in the Deffuant, HK, and ED models. However, this performance consistently declines with increasing noise as the underlying opinion dynamics become harder to recover. In these high-noise settings the optimizer tends to converge toward an explained variance of zero. This again suggests that the optimizer is learning to mimic the null model---and ``freeze" in situations where the signal becomes unclear. 

In the case of the Duggins model, this effect is present even without added noise. As discussed, the model's inherent stochasticity makes it effectively irreproducible, and the optimizer treats it as a high-noise environment from the beginning.

% Interestingly, the Duggins model exhibits a different pattern. The results in Fig. \ref{fig:noise} suggest that, at higher noise levels, the optimizer's performance---relative to the null model baseline---is actually better than in low-noise environments. We hypothesize that this may be due to the model's inherent stochasticity and its ability to approximate the noise addition more flexibly than more deterministic models. This result warrants further investigation.

%These results complete our simulated validation. We now are ready to turn to real-world data to assess model performance in practical settings.

\section{Experiment 4: Real-World Validation}

The previous sections demonstrate that, given an opinion dataset generated by one of the tested models, we can accurately reconstruct the model's parameters. We now test whether this capability extends to real-world data. The underlying assumption is that, if real-world dynamics can be described by one of the models, our optimization method will identify the appropriate model and its parameters. %This validation of opinion dynamics models on empirical data is essential for advancing our understanding of how opinions form and evolve in practice.

% In this section, we describe the dataset used, the specific variables selected, and the performance of our four models in predicting observed patterns in the data.

\subsection{The Dataset: European Social Survey (ESS)}
% \footnote{We received the comment ``The discussion about the dataset used to inform the calibration of the four models should be right after the introduction". While we appreciate this idea and have tried to restructure, we find that it breaks the logical flow. We aim to separate the specific details regarding the simulated data experiments and the real-world experiments, and therefore have not amended this at this time.}}

We used the European Social Survey (ESS)~\citep{ESS2024} as the primary data source. The ESS is an academically driven, cross-national survey conducted biannually since 2002. It contains hundreds of social, political, and attitudinal variables collected across over 30 countries. At the time of writing, eleven rounds of data are available, covering the period from 2002 to 2024.

\subsubsection{Variable Selection:}

We restricted our analysis to the eleven countries that followed the standard ESS protocol across all eleven rounds.\footnote{Although fifteen countries have complete data, due to COVID, four conducted Round 10 independently. These four were excluded for consistency.}

All variables were first normalized to the range [0, 1]\footnote{The notion that opinions can be represented adequately by a number between 0 and 1 has been heavily questioned~\citep{edmonds2005assessing}. However, the models that we test operate under the assumption that the opinions are inputted as scalar values. We support the idea that opinions are much more complicated than this and should be represented as such in future models.}, after which we computed the opinion drift of each dataset, $\Delta_{\text{op}}(D)$. We found that many variables exhibited minimal change between rounds and were therefore deemed uninformative for our purposes. To identify variables with meaningful temporal variation, we selected only those with a $\Delta_{\text{op}}(D) \geq 0.3$. From this filtered set, we selected six variables for analysis.

A notable outcome of this filtering process was that most of the high-variation variables were political in nature, reflecting shifts in public opinion over time. Only four non-political variables passed the 0.3 threshold: \texttt{ppltrst} (trust in people) from Hungary, and \texttt{ppltrst}, \texttt{happiness} (level of happiness), and \texttt{rlgdr} (religious belief level) from Portugal. 

For our non-political subset, we selected \texttt{ppltrst} from Hungary, and \texttt{happy} and \texttt{rlgdr} from Portugal. To complement this, we chose three political variables for comparison: \texttt{imwbcnt} (attitudes toward immigration) from the UK, \texttt{trstun} (trust in the United Nations) from Slovenia, and \texttt{stfdem} (satisfaction with democracy) from Finland.

\subsection{Empirical Results}

\begin{figure}[h!]
    \centering
    \includegraphics[width=0.90\textwidth]{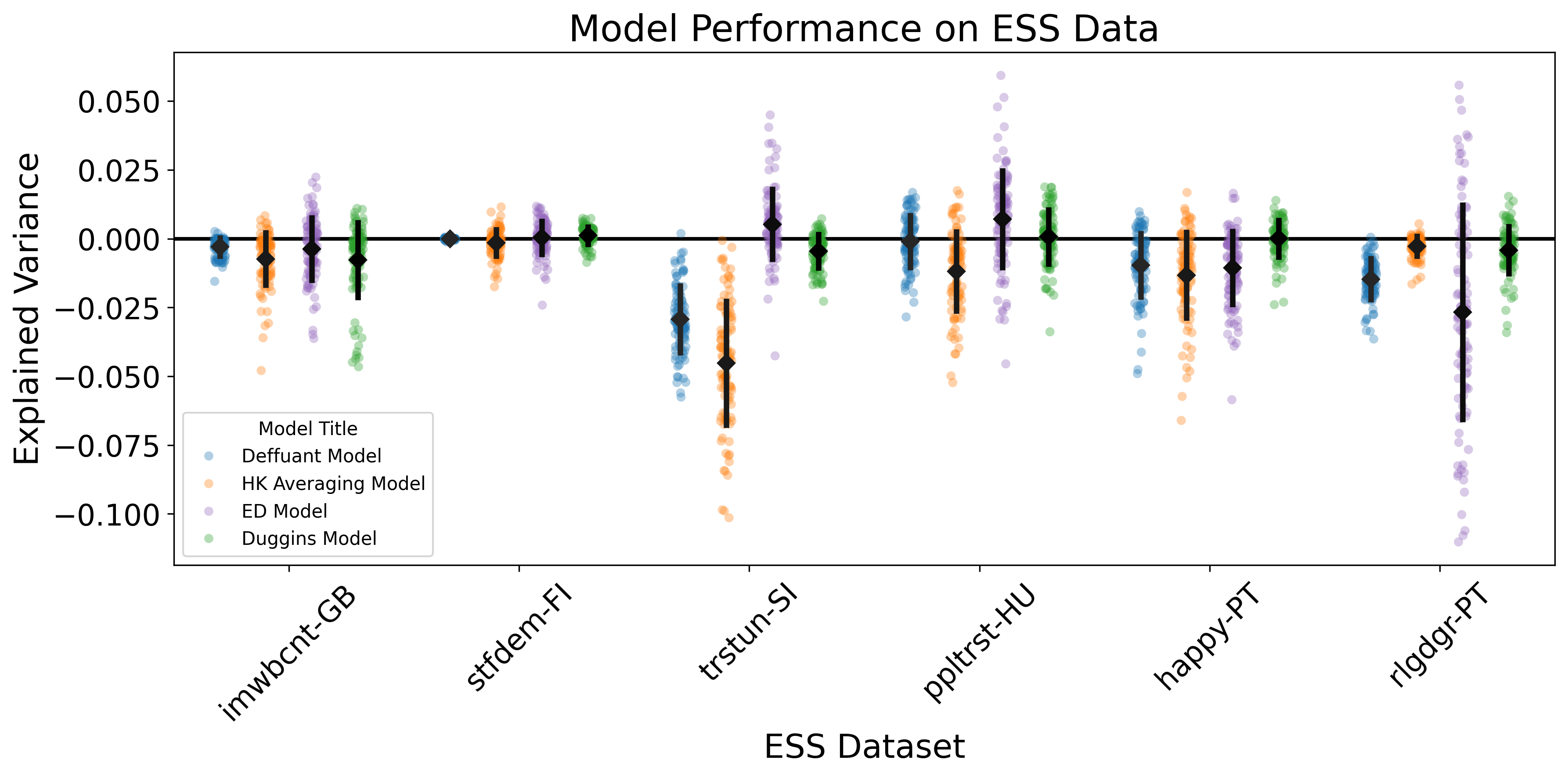}
    \caption{Strip plot showing explained variance for the four models across six ESS datasets. Each strip reflects the distribution of explained variance over multiple runs. The black diamonds and bars are the mean and standard deviation of the runs. Across datasets, models rarely outperform the null model, with explained variance clustering around zero—suggesting the optimizer is learning to ``freeze."}
    \label{fig:model_comparison}
\end{figure}

Fig. \ref{fig:model_comparison} shows the explained variance of each model across the six selected datasets. In each case, the explained variance oscillates around zero. Inspecting the parameters, we again notice that the optimizer tends to converge to parameters that effectively ``freeze" the model. This indicates that the optimizer has identified the null model as the best achievable outcome under the given conditions.

The learned parameter values reveal strategies similar to those observed in high-noise scenarios discussed earlier, suggesting that the optimizer is responding to a similarly low signal-to-noise ratio in the real-world data.

Overall, no model outperforms the null model consistently, suggesting that in these empirical cases, the signal of opinion change is either obscured by noise or directly follows a different model's dynamics.

% \begin{figure}[h!]
%     \centering
%     \includegraphics[width=0.8\textwidth]{figures/real/stripplot.png}
%     \caption{Strip plot showing explained variance for the four models across six ESS datasets. Each strip reflects the distribution of explained variance over multiple runs. The black diamonds and bars represent the mean and standard deviation of the data. Across datasets, models rarely outperform the null model, with explained variance clustering around zero—suggesting the optimizer is learning to ``freeze."}
%     \label{fig:model_comparison}
% \end{figure}

\section{Conclusion and Future Work}

In our analysis, we tested various models of opinion dynamics against real-world data to see if any of them would be able to make accurate predictions. To achieve this, we selected both stylized and more realistic models, as well as a variety of topics for the European Social Survey.

By generating ``synthetic" data using the same model, we were able to get reliable results even with some level of noise. This means that if the dynamics of the data are similar to the one produced by the model, our procedure of calibration and validation would succeed (thus producing high values of explained variance). On the contrary, if the dynamics of data do not follow the model, the validation would fail (i.e. explained variance close to zero).

This study also revealed a limitation of our procedure, as the chosen metrics favour ``reproducible" models. This notion of ``reproducibility" refers to models that tends to produce the same patterns given the same input and the same set of parameters. On the contrary, models which are much more stochastic in their output tend to perform worse even when the data are generated by the same model. This is a criticality of the chosen procedure, which implicitly assumes an almost deterministic approach. However, even the physical world shows multiple cases where multiple outcomes are possible (from quantum mechanics~\citep{griffiths_introduction_2018} to classical probability~\citep{feller1968}). Thus, we hope that future works will improve our approach, by producing a methodology still open to multiple outcomes with different probabilities.

Despite this issue, all models seemed to perform at best like our null model, which just supposes that the future state of the system is equal to the previous state. Indeed, all models in their optimization phase learned to ``freeze" in order to mimic the null model. We evaluated ESS data with a significant amount of opinion drift ($\geq 0.3)$---ensuring that the ``freezing" is not due weak signals from limited opinion change. We show in Section~\ref{sec:validating_optimizer} that under this level of opinion drift, the optimizer should be able to capture the underlying dynamics, given that the dynamics are within the model's representational capacity. Therefore, the poor results of the models are not simply due to their lack of reproducibility, but mainly due to their inability to represent the dynamic sequences of observed opinion evolutions solely with an adjustment of the free parameters\footnote{In mathematical terms, this implies that the model does not have the representational capacity to recreate the observed opinion evolution. In order to represent the ground-truth opinions, a different model must be used.}. Alternatively, this can be phrased as the fact that the dynamics of real-world opinion distributions do not seem to follow any of the dynamics presented in the tested models.

Of course, these results should not be over-generalized to all opinion dynamics models. Indeed, the current test was limited to four opinion dynamics models, even if it included one model derived from experimental data and thus, whose micro-dynamics was already validated~\citep{carpentras2022}. However, the current test suggests that the literature on opinion dynamics still needs to work heavily to develop models that can reproduce real-world opinion dynamics, and thus be employed for real-world applications such as policy design. We hope that, in the future, more attention will be dedicated to testing and improving models' validity.

\section*{Acknowledgements}

The authors are grateful for support by the project “CoCi: Co-Evolving City Life”, which received funding from the European Research Council (ERC) under the European Union’s Horizon 2020 research and innovation programme under grant agreement No. 833168. We would also like to thank the reviewers of the Social Simulation Conference (SSC25) for their constructive critiques and helpful encouragement, both of which helped refine and strengthen this manuscript.
% \begin{thebibliography}{6}

% \end{thebibliography}

\bibliography{main}
\bibliographystyle{styles/bibtex/spbasic_unsrt}

\end{document}